\documentclass[]{spie}  %>>> use for US letter paper
\usepackage[]{graphicx}

\title{The focal plane instrumentation for the DUNE mission} 

\author{J. Booth\supit{a}, M. Cropper\supit{b}, F. Eisenhauer\supit{c},   A. Refregier\supit{d}   \& the DUNE collaboration\supit{e}
\skiplinehalf
\supit{a}{\small Jet Propulsion Laboratory, California Institute of Technology, Pasadena, CA, USA 91109}\\
\supit{b}{\small Mullard Space Science Laboratory, University College London, Dorking, Surrey RH5 6NT, UK}\\
\supit{c}{\small Max Planck Institute for Extraterrestrial Physics, Postfach 1312, 85741 Garching, Germany}\\
\supit{d}{\small Service d'Astrophysique, CEA Saclay, F-91191 Gif sur Yvette, France}}

\authorinfo{\supit{c}{\bf DUNE collaboration:} {\bf Co-investigators:} A. Refregier ({\bf PI}, CEA Saclay) 
M. Douspis (IAS Orsay) Y. Mellier (IAP Paris) 
B. Milliard (LAM Marseille)
P. Schneider (U. Bonn) H.-W. Rix (MPIA) 
R. Bender (MPE Garching) F. Eisenhauer (MPE Garching) R. Scaramella (INAF-OARM)
% ...
\\ {\bf Authors'
  contacts:} Jeffrey.T.Booth@jpl.nasa.gov}

\begin{document} 
  \maketitle

%%%%%%%%%%%%%%%%%%%%%%%%%%%%%%%%%%%%%%%%%%%%%%%%%%%%%%%%%%%%% 
\begin{abstract}
DUNE (Dark Universe Explorer) is a proposed mission to measure
parameters of dark energy using weak gravitational lensing The
particular challenges of both optical and infrared focal planes and
the DUNE baseline solution is discussed. The DUNE visible Focal Plane
Array (VFP) consists of 36 large format red-sensitive CCDs, arranged
in a 9x4 array together with the associated mechanical support
structure and electronics processing chains. Four additional CCDs
dedicated to attitude control measurements are located at the edge of
the array. All CCDs are 4096 pixel red-enhanced e2v CCD203-82 devices
with square 12 $\mu$m pixels, operating from 550-920nm. Combining four
rows of CCDs provides a total exposure time of 1500s. The VFP will be
used in a closed-loop system by the spacecraft, which operates in a
drift scan mode, in order to synchronize the scan and readout
rates. The Near Infrared (NIR) FPA consists of a 5 x 12 mosaic of 60
Hawaii 2RG detector arrays from Teledyne, NIR bandpass filters for the
wavelength bands Y, J, and H, the mechanical support structure, and
the detector readout and signal processing electronics. The
FPA is operated at a maximum temperature of 140 K for low dark current
of 0.02e$-$/s. Each sensor chip assembly has 2048 x 2048 square pixels
of 18 $\mu$m size (0.15 arcsec), sensitive in the 0.8 to 1.7 $\mu$m wavelength
range. As the spacecraft is scanning the sky, the image motion on the
NIR FPA is stabilized by a de-scanning mirror during the integration
time of 300 s per detector. The total integration time of 1500 seconds
is split among the three NIR wavelengths bands. DUNE has been proposed
to ESA's Cosmic Vision program and has been jointly selected with
SPACE for an ESA Assessment Phase which has led to the joint Euclid
mission concept.

\end{abstract}

%>>>> Include a list of keywords after the abstract 


%%%%%%%%%%%%%%%%%%%%%%%%%%%%%%%%%%%%%%%%%%%%%%%%%%%%%%%%%%%%%
\section{INTRODUCTION}
\label{sec:intro}  % \label{} allows reference to this section

The Dark Universe Explorer (DUNE) is a proposed wide-field space
imager whose primary goal is the study of dark energy and dark matter
with unprecedented precision using weak gravitational
lensing. Immediate secondary goals focus on the evolution of galaxies,
to be studied with unprecedented statistical power, the detailed
structure of the Milky Way and nearby galaxies, and the demographics
of Earth-mass planets. DUNE is a medium-class mission as defined by
ESA’s Cosmic Vision program, consisting of a 1.2m Korsch-like
three-mirror telescope with a combined visible/near-infrared
field-of-view of 1 square degree and 0.23 arcsec resolution (visible)
in a geosynchronous orbit. DUNE plans to carry out an all-sky survey
in one visible and three NIR bands which will form a unique legacy for
astronomy. DUNE thus addresses multiple goals of the ESA Cosmic Vision
program and would yield major advances in a broad range of fields in
astrophysics and cosmology (http://www.dune-mission.net). DUNE is a
realization of the wide-field imaging mission recommended by the
ESO/ESA Working Group on Fundamental Cosmology (WGFC).  

One of the most powerful tools to tackle dark energy is weak
gravitational lensing of background galaxies by foreground dark
matter: this forms the core of the DUNE
mission\cite{ref08,refrdous}. Gravitational deflection of light by
intervening dark matter concentrations causes the images of background
galaxies to acquire an additional ellipticity on the order of one
percent. Utilization of this cosmological probe relies on the
measurement of image shapes and redshifts, or distances, for several
billion galaxies, both requiring space observations for point spread
function (PSF) stablity and photometric measurements over a wide
wavelength range in the visible and especially near-infrared
(NIR). These are the driving requirements on the instrumentation of
DUNE, and are solved by the separate, large visible and NIR focal
plane arrays which will be discussed in more detail in this paper.

\subsection*{Mission and orbital requirements}

The driving requirements for the mission design of DUNE are from the
wide extragalactic survey and specifically the need for the stability
of the PSF and the large coverage of the sky. PSF stability puts
stringent requirements on pointing and thermal stability during the
observation time. The 20,000 square degrees survey demands high
operational efficiency, which can generally be achieved either in a
step-and-stare imaging mode or by continuously scanning the sky during
science acquisition. The pointing accuracy requirement is identical in
both cases, namely 0.2 $\mu$rad over 375 s integration time per CCD chip
(to reach magnitude limits required by science goals), and is
essentially driven by the elementary integration time and the spatial
resolution. This can be achieved using a drift scanning mode (or Time
Delay Integration, TDI, mode) for the CCDs in the visible focal plane,
as for GAIA. A continuous scanning mode is by nature more efficient
than the step-and-stare mode, since the science measurements can be
continuously achieved over a relatively long fraction of a great
circle, which reduces the number of maneuvers by one or two orders of
magnitude, and is selected as the baseline for DUNE.  For the infrared
focal plane, using available HgCdTe arrays, TDI mode not feasible and
the proposed concept is to stop the image motion during the
integration time in the NIR focal plane by using a small de-scan
mirror located close to the telescope exit pupil. In practice, the
de-scan design is such that this mirror is operated in quasi-static
mode with low amplitude oscillations, about 1 degree and a period of
300 seconds.  

\section{DUNE Payload and Focal Plane Instrumentation}

The Payload module (PLM) design uses Silicon Carbide (SiC) technology
for the telescope optics and structure. This provides low mass, high
stability, low sensitivity to radiation and the ability to operate the
entire instrument at cold temperature, typically below 170 K, which
will be very useful for cooling the large focal planes. The telescope
structure supports the Focal Plane Assemblies.  Located behind the
passively cooled primary mirror are the two offset focal planes
(Visible and NIR), each about 250x500mm and 0.5 deg2 field of
view. The visible focal plane is populated by 36 4096x4096 CCD e2V
203-82 (with 4 additional CCDs for use by the attitude control system
of the spacecraft), and the NIR focal plane has 60 2048x2048 Teledyne
HAWAII-2RG HgCdTe arrays mosaiced together. Defined as part of the
payload are also the de-scan mirror mechanism and the additional
payload data handling unit (PDHU). Key payload parameters are listed
in Table 1.

\begin{table}
\begin{center}
\caption{Key payload parameters}
\label{baseline}
\begin{tabular}{|ll|}
\hline
\bf Instrument& \bf Value/description \\\hline
Optical configuration&Off-axis Three Mirror Anastigmat (TMA)\\\hline
Pupil diameter&1.2m\\\hline
spectral range& VIS: 500-900nm\\
              & NIR: 920-1600nm (3bands) \\ \hline
Resolution& 0.102 arcsec\\
(pixel size)&0.15 arcsec\\\hline
Visible FOV& $1.08^\circ \times 0.48^\circ$ \\
&(array of 9x4 matrices / 4096x4096 pixels per matrix / 12$\mu$m pitch) \\
NIR FOV& $1.04^\circ \times 0.44^\circ$ \\
&(array of 5x12 matrices / 2048x2048 pixels per matrix / 12$\mu$m pitch) \\  \hline\hline
\bf Focal plane \& observing mode & \\\hline
Visible Channel& TDI through S/C slow rotation \\
&(1500s integration time per celestial object)\\
NIR channel& Step\&Stare through de--scan mechanism \\
&(600s integration time per celestial object for J, H bands, 300s for Y band)\\\hline\hline
\bf Focal planes Mass \& Power & \\  \hline
Visible + NIR FPA + electronics mass& 155kg (CBE)\\
Visible + NIR FPA + electronics power& 369W (CBE)\\
\hline
\end{tabular}

\end{center}
\end{table}

\subsection{Telescope} 

The optical concept is a Korsch-like f/20 three-mirror telescope
(Figure 1). The third mirror is slightly off-axis to separate the two
channels: visible and NIR. After the first two mirrors, the optical
bundle is folded just after passing the primary mirror (M1) to reach
the off-axis tertiary mirror. Then, a dichroic element located near
the exit pupil of the system provides the spectral separation of the
visible and NIR channels. On the NIR optical path, the implementation
of the de-scan mechanism close to the dichroic filter allows for a
largely symmetric configuration of both spectral channels. After a
final folding, the bundles are directed towards the focal planes. This
dichroic configuration minimizes the field of view necessary to
implement the two large focal planes, leading to a rather compact
system.

\subsection{Visible Focal Plane Array}

The baseline for the visible Focal Plane Array (VFP) consists of 36
large format red-sensitive CCDs, arranged in a 9x4 array (Figure 2)
together with the associated mechanical support structure and
electronics processing chains. Four additional CCDs are dedicated to
the attitude and orbit control system (AOCS) measurements are located
at the edge of the array. All CCDs are 4096 pixel red-enhanced e2v
CCD203-82 devices with square 12 $\mu$m pixels. The CCDs are standard
devices requiring no development beyond a rerouting of the connections
to eliminate one of the two flexi-lead connections, since only one
video output per CCD is required. No development for additional
features is envisaged. They are 4-sides buttable with a dead space of
1.6mm between active areas, providing a filling factor of 94%. The
physical size of the array is 466x233 mm corresponding to
1.09$^\circ$x0.52$^\circ$. Each pixel is 0.102 arcsec, so that the PSF is well
sampled in each direction over approximately 2.2 pixels.  

\begin{figure}[!h]
\begin{center}
\begin{tabular}{c}
\includegraphics[width = 0.75\textwidth, angle=0]{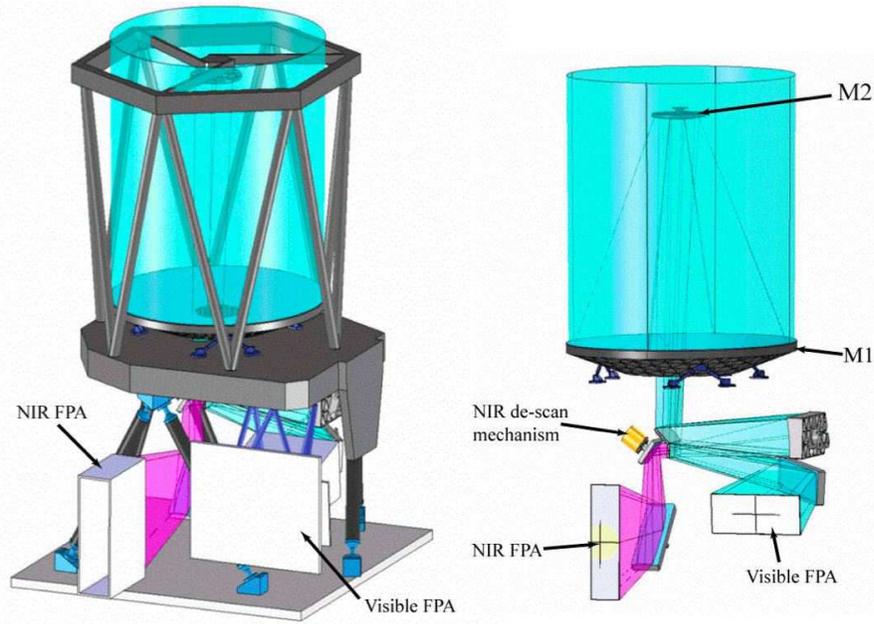}
\end{tabular}
\end{center}
\caption{A side view of the DUNE telescope showing the location of the
main mirror elements and the focal plane assemblies.  }
\label{fig:instrument}
\end{figure}

The VFP operates in the red band from 550-920nm. This bandpass is
produced by the dichroic. If further filtering is required, an optical
filter (long-pass such as the Schott GG495) will be mounted near the
CCDs. Care will need to be taken with ghost images (which are very
important in the case of DUNE), and filters for each CCD will be
mounted individually on a carrier to optimize glass thickness. These
filters will provide some additional shielding for the CCDs. An
alternative being considered in the assessment phase is to add a
second filter coating on the optical channel fold mirror.  

The optical camera operates in TDI mode, so that the CCD rows are
continuously clocked from a single readout node at the same rate as
the satellite scans the sky (1.12 arcsec/sec). The CCDs are 4-phase
devices, so they can be clocked in 1⁄4 pixel steps. The exposure
duration on each CCD is 375s, permitting a slow readout rate and
minimizing readout noise. Combining 4 rows of CCDs will then integrate
for a total time of 1500s. With the broad bandpass this ensures good
sensitivity.

The required readout rate from the CCDs is slow, which is advantageous
for minimizing system noise. Each CCD will have a dedicated processing
electronics module (PEM) responsible for reading out the CCD, bias
generation, signal conditioning and digitization. Digitization will be
to 16-bit. Each row of 9 processing modules will in turn connect to a
common power supply and data routing module, which will transmit the
data to the onboard processing for compression and packetization.

The structure of the VFP will provide a stable optomechanical support
for the detectors and their processing chains as shown in Figure
2. These will be passively held at 170K to minimize the CTE effects
from radiation damage. This temperature also reduces the total heating
budget for the payload. The structure material is to be determined and
will be selected after a trade-off with due regard to the interface
conditions with the optical bench and the CCD package, the
thermo-mechanical stability, the ease and cost of manufacture and the
cleanliness constraints. The electronics will operate in a warm
environment, thermally decoupled from the CCDs and the rest of the
payload. A radiator will be used to dump the dissipation from the CCDs
(and parasitic heat loads) to space, while the structure housing the
electronics will act as their radiator. There are no mechanisms in the
VFP. 

With a spatial resolution of the PSF in each direction of 2.2 pixels
(FWHM), the contribution to the spatial degradation induced by the
detector MTF is small, even at far red wavelengths. Additional sources
of error including TDI rate errors and the broadening owing to the 1/4
pixel step size during TDI clocking will also be small. The overall
sampling including all of these contributions is optimal at
2.2. pixels per FWHM. The effect on image quality as a result of
radiation damage will need to be quantified and will draw on heritage
from the GAIA program. 

The VFP will be used by the spacecraft in a closed-loop system to
ensure that the scan rate and TDI clocking are synchronized. The two
pairs of AOCS CCDs (see Figure 2) provide two speed measurements on
relatively bright stars (V magnitude 22-23, a few thousand total
collected photo-electrons). This technique will have been proven in
the framework of GAIA, and similar principles will be followed for
DUNE, where the requirements are for a slower scan rate but very
similar pointing accuracy.  

The DUNE VFP is largely a self-calibrating instrument. For the shape
measurements, stars of the appropriate magnitude will allow the PSF to
be monitored for each CCD including the effects of optical distortion
and detector alignment. The PSF will be calibrated as a function of
wavelength pre-launch, as it is wavelength dependent, and will differ
for different spectral energy distributions. Radiation-induced charge
transfer inefficiency will modify the PSF and will need to be
calibrated both through the in-orbit self-calibration, and also
extensively in the on-ground calibration at CCD level. This effect is
being exhaustively investigated in the GAIA program, where centroiding
shifts of 0.001 pixel have to be measured. DUNE is mainly sensitive to
PSF shape variations and modelling of the effects will be required in
order to achieve a detailed understanding and a validation of the
overall ground calibration procedure. For throughput measurements,
there will be sufficient transits of stars of known magnitude to
calibrate spatial variations both from the optics and the CCDs. Other
calibrations will include scattered light and detector characteristics
(noise levels, cosmetics).  

The DUNE VFP uses standard technology with essentially qualified
detectors. Most of the conceptual design can be adapted from other
missions, in particular GAIA. Radiation damage to the detectors is one
issue in DUNE that will need careful attention, as the shape
measurements have to be made accurately. However, software
post-processing and in-flight calibration techniques are being
developed to address this issue.

\begin{figure}
\begin{center}
\begin{tabular}{c}
\includegraphics[width = 0.7\textwidth, angle=0]{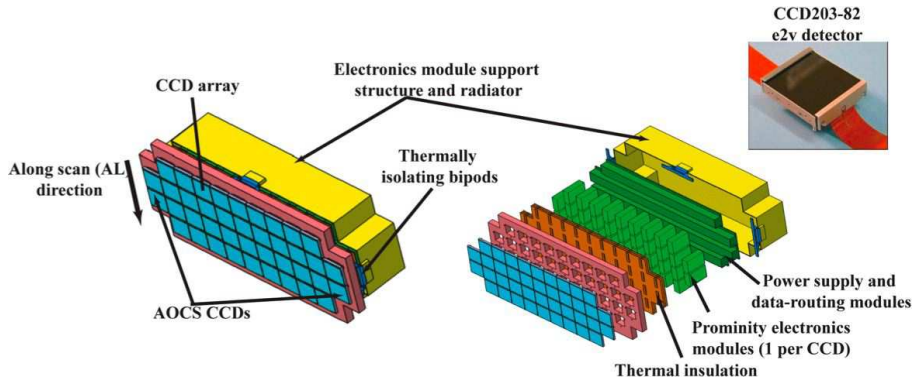}
\end{tabular}
\end{center}
\caption{Left: The VFP assembly with the 9x4 array of CCDs and the 4
AOCS sensors on the front (blue) and the warm electronics radiator at
the back (yellow). Right: An expanded view of the VFP assembly,
including the electronics modules and thermal hardware (but excluding
the CCD radiator). Inset: The e2v CCD203-82 4kx4k pixels shown here in
a metal pack with flexi-leads for electrical connections. One of the
flexi-leads will be removed.}
\label{fig:2}
\end{figure}

\subsection{Near Infrared Focal Plane Array}

The NIR system meets the science requirements and is based on mature
components with which JPL, MPE and MPIA have had extensive experience
in characterizing and integrating into astrophysical instruments. The
baseline NIR FPA is composed of a 5 x 12 mosaic of 60 Hawaii 2RG
detector arrays from Teledyne, NIR bandpass filters for the wavelength
bands Y, J, and H, the mechanical support structure, and the detector
readout and signal processing electronics. The FPA is
operated at a maximum temperature of 140 K for low dark current of
0.02e-/s\footnote{M.Rao (Teledyne), private communications.}.  Each
array has 2048 x 2048 square pixels of 18 $\mu$m size resulting in a
0.15 x 0.15 arcsec2 field of view (FOV) per pixel for the image scale
of DUNE. The 5 x 12 mosaic has a physical size of 482 x 212 mm, and
covers a FOV of 1.04$^\circ$ x 0.44$^\circ$ or 0.46 square degrees with a filling
factor of 89% due to gaps between the arrays. The FPA structure is
made from molybdenum to match the thermal expansion coefficient of the
array mounts. The total weight of the FPA, including the Mo-structure
designed for loads of 25g in all axes, is around 40 kg without
light-weighting. With the expected availability of a SiC array
mounting option from Teledyne, the complete FPA structure can likewise
be made from SiC to reduce the weight further.

The HgCdTe Hawaii 2RG arrays are standard devices requiring no
development with flight instrument heritage (WISE and James Webb Space
Telescope (JWST)). They are composed of a CMOS multiplexer bump-bonded
to the HgCdTe sensor layer which is sensitive in the 0.8 to 1.7 μm
wavelength range with a quantum efficiency of about 0.8. Unlike the
charge transfer process used by the CCD, the CMOS multiplexer
individually addresses, non-destructively reads, and resets each
pixel. Up to 32 data channels are available for parallel output of
video data. The array can be read at a maximum pixel rate of 5MHz per
channel, but for low read noise operation, the pixel rate is typically
limited to 100kHz, resulting in 1.3 seconds to read out the entire
array.  

For each array, the readout control, A/D conversion of the video
output, and transfer of the digital data via a serial link is handled
by the SIDECAR ASIC (application specific integrated circuit)
developed by Teledyne for JWST. This device has low power consumption
(about 100 mW for 100kHz pixel rate and 16-bit A/D-conversion of 32
channels), high flexibility (programmable readout, including
windowing), and an operating temperature range from 30 to 300 K. The
ASICs will be located within a few centimetres from the arrays to
avoid degradation of the analogue video signals, but will be thermally
isolated from the arrays and can thus be operated at a higher
temperature.

\begin{figure}
\begin{center}
\begin{tabular}{c}
\includegraphics[width = 0.7\textwidth, angle=0]{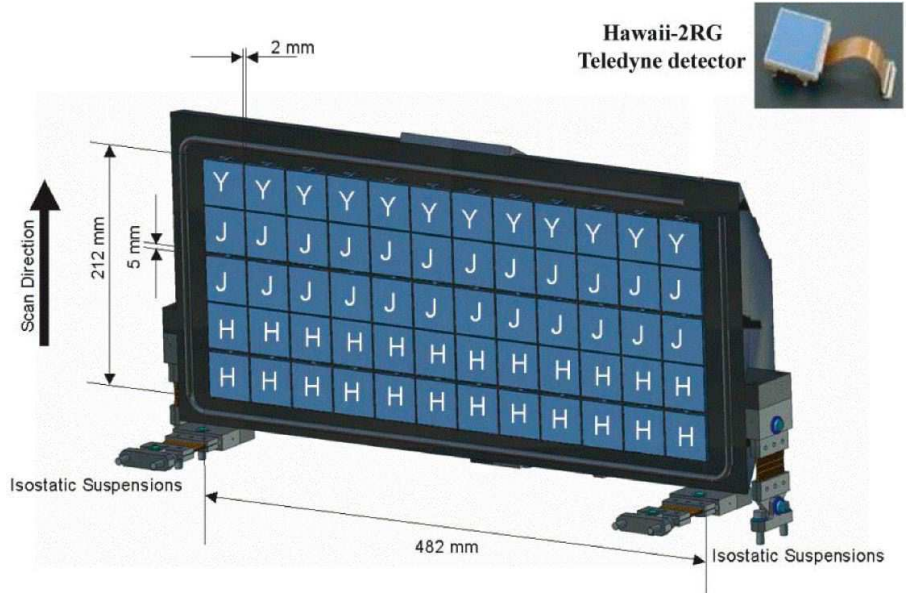}
\end{tabular}
\end{center}
\caption{Layout of the NIR FPA (MPE/Kayser-Threde). The 5 x 12 Hawaii
2RG Teledyne detector arrays (shown in the inset) are installed in a
molybdenum structure}
\label{fig:3}
\end{figure}

As the spacecraft is scanning the sky along a great circle, the image
motion on the NIR FPA is stabilized by a de-scanning mirror during the
integration time of 300s or less per NIR detector. After the initial
integration, the mirror flips back to its start position, thus
shifting the field from one row of the array to the following
exposure. In this way, the total integration time of 1500 s for the
0.4$^\circ$ high field is split among five rows and 3 wavelengths bands along
the scan direction. Since two arrays are covered with J and H filters
and the other one with a Y band filter, the effective integration
times are 600 s in J and H, and 300 s in Y. To achieve the limiting
magnitudes defined by the science requirements (24AB, SNR=5, point
source) within these integration times, an effective read noise of
about 5 e- is required (which would achieve better than 24.4 AB). The
read noise of the arrays is about 18 e- for a single read (25 e- for
correlated double sampling). For n reads, the effective read noise
goes down roughly as the square root of n, implying that a minimum of
13 reads are required to meet the sensitivity specifications. Given
that one read takes 1.3s, many readouts are feasible. The number of
reads is a trade-off between sensitivity on one side and available
computing power and power consumption on the other side. The
processing of the digital video signals requires more computing power
and memory than provided by the ASICs. Therefore, data are processed
on a dedicated unit located in the service module. Images can be
'integrated up the ramp' and stored in 32 bit format in a 1GB
memory. For buffering data, another GB may be required. An average
processing power of 22 MFLOPS is sufficient to process 13 reads within
300 s. For a larger number of reads, the computing power increases
with the number of reads; for permanent readout, 390 MFLOPS are
required. For rejection of cosmic ray events and data compression,
additional computing power and storage is required.  

No specific hardware is required in the NIR FPA for pointing
control. The spacecraft will control all attitude and pointing for the
NIR channel, including the de-scanning mirror, by using the CCD
detectors located in the visible focal plane and dedicated to the
AOCS.  

The NIR FPA will operate in a step and stare operation with the
de-scanning mirror. This is the only science mode. Other modes will be
required during acquisition and scan-rate locking. The NIR FPA readout
will be operated in one of two modes to be optimised during the
assessment phase: either a “sample up the ramp” mode, or multiple
Fowler sampling. “Up the ramp” means evenly spaced in time
non-destructive reads will be made during each integration
period. Multiple Fowler sampling clusters a set of reads at the
beginning and end of integration times. Additional on-board processing
will be required, beyond the SIDECARs, to perform either of these
sampling schemes. The primary calibration requirement will be for
occasional dark current status. Dark frames will be taken by using the
scanning mirror to direct light away from the NIR FPA, effectively
acting as like a shutter would for a dark frame.  

The DUNE NIR FPA uses standard technology with qualified detectors and
electronics. Most of the conceptual design can be adapted from other
missions, in particular JWST. Though a baseline design has been
identified, future trades will be studied, including alternative
HgCdTe detector providers both European (eg. QinetiQ and Sofradir) as
well as from the United States (eg. Raytheon), as well as the
possibility of using InGaAs detectors on the Hawaii 2RG multiplexer
with a 1.6$\mu$m cutoff. Additional studies will be conducted to
determine the optimal method for efficient use of the NIR in the drift
scan mode of the mission. These trades include options of controlled
scanning of the NIR FPA itself rather than a counter scanning mirror,
and fast readout modes of the detector to match the drift rate in a
pseudo-TDI mode. Persistency effects on the arrays may be the most
critical issues. Tests are already underway under realistic conditions
(operating temperature and background cutoff wavelength) to quantify
and manage the effect.

\subsection{Technological readiness}

Table 2 shows the technological readiness (TRL) of the different DUNE
components, along with their heritage status. As can be seen in this
table, the mission has a low level of technological risk because of
the use of only ‘off-the-shelf’ components and its reliance on the
heritage.

While the level of technological risk of DUNE is low, several
technological and system studies could be performed early to minimize
scheduling risks. In particular, a study of the impact of radiation on
the charge transfer efficiency of CCDs and the performance of these
detectors at low flux levels should be carried out early as it may
impact the orbit choice. The performance of the NIR HgCdTe detectors,
whose image quality performance (persistence, pixel cross talk, etc.)
needs to be further quantified and its mode of operation optimized
(eg. number of multiple reads) The performance of a-posteriori PSF
calibration and its impact on shape measurements also needs to be
further studied to better specify the requirements on the AOCS system
and thermo-mechanical perturbations.

While the VFP is a large array, all aspects of the optical focal plane
have significant heritage through other ESA programs, in particular
GAIA, which has an FPA three times larger. The e2V CCD203-82 detectors
have been qualified by e2v for Lockheed Martin for the Solar Dynamics
Observatory (SDO) to be launched at the end of 2008. The technological
readiness for the CCD proximity electronics chain and structure has
been developed both at Mullard Space Science Laboratory (MSSL) and
Rutherford Appleton Laboratory (RAL) for GAIA and Solar Dynamic
Observatory (SDO), and these have already demonstrated the required
performances (linearity, noise and radiation hardness). Although NIR
FPA is a large array which will present specific challenges, all
aspects of the NIR focal plane have significant heritage (TRL 6)
through other space projects on a smaller scale, in particular NASA’s
JWST.

\begin{table}
\begin{center}
\caption{Technnology readiness levels for key payload subsystems}
\label{baseline}
\begin{tabular}{|l|l|l|}
\hline

Subsystem&
Heritage status&
TRL\\ \hline
Optics&
New design derived from Aladin, Gaia, RocSat2 or THEOS&
7\\
Filters&
New design&
7\\
Dichroic&
New design&
6\\
Structure&
Gaia&
7\\
Thermal control&
Standard equipment&
7\\
De-scan mirror&
LOLA fine pointing mirror&
5\\
CCDs&
Minor modifications to design used for NASA-SDO&
6\\
CCD electronic chain &
Gaia/Eddington &
6\\
VFPA structure&
Herschel/Gaia/Eddington&
6\\
NIR Detector HgCdTe&
HST WFC3, JWST NIRCAM and NIRSpec, WISE&
6\\
Hawaii 2RG multiplexer&
HST WFC3, JWST NIRCAM and NIRSpec, WISE&
6\\
NIR electronic chain&
HST WFC3, JWST NIRCAM and NIRSpec&
6\\
NIRFPA structure&
Herschel/Gaia&
6\\
\hline
\end{tabular}

\end{center}
\end{table}

\subsection{Instrumentation evolution for the Euclid mission}

DUNE has been proposed to ESA’s Cosmic Vision program and has been
jointly selected with SPACE\cite{space} for an ESA Assessment Phase
which has led to the joint Euclid mission concept. Euclid extends the
DUNE science methodology to include another dark energy probe (baryon
acoustic oscillations (BAO)). The expanded science and instrumentation
will have significant benefits and impacts on primary and secondary
science, as well as payload and mission design. Euclid will have two
primary instruments (an imaging channel with visible and NIR FPAs for
weak lensing, and a NIR multi-object spectroscopic channel for BAO),
with three separate focal planes. Mission architecture changes include
utilizing an L2 orbit, and (probably) a step-and-stare observing
mode. The photometric redshift channel (the NIR FPA for DUNE) would be
reduced in size compared to the original in DUNE to between 12 and 16
detectors total, with a corresponding plate scale change per pixel to
still cover 0.5 square degrees, and a descan mechanism would not be
required. Significant effort is ongoing in continuing design trades
for the combination of all three focal planes for maximum science
return while minimizing cost and risk.

%%%%%%%%%%%%%%%%%%%%%%%%%%%%%%%%%%%%%%%%%%%%%%%%%%%%%%%%%%%%%
\acknowledgments     %>>>> equivalent to \section*{ACKNOWLEDGMENTS}       

This work was performed in part at The Jet Propulsion Laboratory,
which is operated by the California Institute of Technology under a
contract from NASA.We thank CNES for support on an earlier version of
the DUNE mission and EADS/Astrium, Alcatel/Alenia Space, as well as
Kayser/Threde for their help in the preparation of the DUNE concept.

%%%%%%%%%%%%%%%%%%%%%%%%%%%%%%%%%%%%%%%%%%%%%%%%%%%%%%%%%%%%%
%%%%% References %%%%%

\bibliography{dune_spie2}   %>>>> bibliography data in report.bib
\bibliographystyle{spiebib}   %>>>> makes bibtex use spiebib.bst

\end{document}